\documentclass[12pt]{article}
\RequirePackage[OT1]{fontenc}
\usepackage[english]{babel}
\usepackage{amsmath}
\usepackage{amssymb}
\usepackage{authblk}
\usepackage{graphicx,psfrag,epsf}
\usepackage{enumerate}
\usepackage{natbib}
\usepackage{threeparttable}
\usepackage{multirow}
\usepackage{stackengine}
\usepackage{setspace}
\usepackage{float}
\usepackage{appendix}
\usepackage{stackengine}
\usepackage{verbatim}
\usepackage[top=1in, bottom=1in, left=1in, right=1in]{geometry}

\usepackage{nameref}
\usepackage{varioref}
\usepackage{url}
\usepackage{comment}
\usepackage{lscape}
\usepackage{cleveref}
\newcommand\independent{\protect\mathpalette{\protect\independenT}{\perp}}
\def\independenT#1#2{\mathrel{\rlap{$#1#2$}\mkern2mu{#1#2}}}

\setstackgap{L}{.7\normalbaselineskip}
\setstackEOL{\#}
\DeclareMathSymbol{\E}{\mathbin}{AMSb}{"45}
\newcommand{\EE}[1]{\E \left [ #1 \right ]}

\newcommand{\blind}{1}

\begin{document}

\def\spacingset#1{\renewcommand{\baselinestretch}%
{#1}\small\normalsize} \spacingset{1}


\if1\blind
{
  \title{\bf
   Bayesian analysis of longitudinal studies with treatment by indication}
  \author[1]{Reagan Mozer\thanks{Corresponding author: Reagan Mozer (rmozer@bentley.edu).}}
  \author[2]{Mark E. Glickman}
  \affil[1]{Bentley University}
  \affil[2]{Harvard University}

  \maketitle
} \fi

\if0\blind
{
  \bigskip
  \bigskip
  \bigskip
  \begin{center}
    {\LARGE\bf Bayesian analysis of longitudinal studies with treatment by indication}
\end{center}
  \medskip
} \fi

\bigskip
\begin{abstract}
In a medical setting, observational studies commonly involve patients who initiate a particular treatment (e.g., medication therapy) and others who do not, and the goal is to draw causal inferences about the effect of treatment on a time-to-event outcome. A difficulty with such studies is that the notion of a treatment initiation time is not well-defined for the control group. We propose an approach to estimate treatment effects in longitudinal observational studies where treatment is given by indication and thereby the exact timing of treatment is only observed for treated units. We present a framework for conceptualizing an underlying randomized experiment in this setting based on separating the process that governs the time of indication for treatment from the mechanism that determines assignment to treatment versus control. Our approach involves inferring the missing indication times followed by estimating treatment effects. This approach allows us to incorporate uncertainty about the missing times of assignment, which induces uncertainty in both the selection of the control group and the measurement of time-to-event outcomes for these controls. We demonstrate our approach to study the effects on mortality of inappropriately prescribing phosphodiesterase type 5 inhibitors (PDE5Is),  a medication contraindicated for certain types of pulmonary hypertension, using data from the Veterans Affairs (VA) health care system.
\end{abstract}

\noindent%
{\it Keywords:} causal inference, observational studies, comparative effectiveness research
\vfill

\newpage
\spacingset{1.45} 
\section{Introduction}
\label{sec:intro}

In certain observational studies, particularly in a medical context, interest centers on estimating the causal effects of an action, treatment, or intervention on a time-to-event outcome (e.g., the effects of surgery on post-operative survival time). Insights about the effects of these non-randomized treatments have the potential to help answer important questions both in population health and in individualized medicine.  In these settings, measuring time-to-event outcomes in a treated group is straightforward; time is measured from treatment initiation. However, a key obstacle for inference is the lack of observed initiation times in the control group, without which a time-to-event outcome is not defined. 

The question of how to make valid causal inferences in this setting using longitudinal, observational data becomes further complicated with the possibility of confounding due to ``treatment by indication" \citep{poses1995controlling}. 
This issue stems from the principle that a good clinician will initiate a new treatment only when evidence exists that the candidate treatment is medically necessary for the patient in question \citep{poses1995controlling}. 
In studies evaluating the effects of a non-randomized treatment on a population with a common disease or condition, this means that treatment is initiated (and, therefore, assignment to treatment is observed) only when the potential benefits outweigh the anticipated risks or side effects associated with treatment. 
As a result, subjects who receive treatment during the study period may differ systematically from untreated subjects observed during the same time period in terms of their prognosis. 
This type of confounding can generally be avoided by comparing groups of subjects who presented with similar indications at similar times after their initial diagnosis; however, these variables are typically not recorded for subjects who do not receive treatment. 
This issue is especially common in studies that rely on data from electronic medical records, administrative databases, and health registries \citep{byar1980data, levine2008registries, freidlin2012assessing}.
Thus, for the group of untreated subjects, it is unclear whether treatment was indicated at an unknown time during the study and was purposefully withheld at that time or if treatment was simply never considered due to lack of indications. 
A primary challenge in this domain therefore concerns how to infer the unknown indication times from the observed data.

In this paper, we consider how to draw causal inferences about a binary treatment, which can either be initiated or withheld from a given subject at a single point in time. This time point, which we refer to as the ``indication time", is the time at which a subject presents with a particular set of symptoms or pre-specified indications that necessitate clinical intervention. The indication time for any individual might be a function of the subject's behavior (e.g., when a subject requests a medication therapy), or might be solely determined by clinical factors. For instance, in our applied example, we consider a population of patients in the VA health system who are diagnosed with certain types of pulmonary hypertension (PH), where individuals may receive medication therapy in the form of phosphodiesterase-5-inhibitors (PDE5Is) for treatment of PH-related symptoms only when it is determined that the medication may be beneficial given the severity of the patient's condition.
As a result, indication times may vary greatly across patients in a sample, which induces a complicated time structure in the observed data. 
To accommodate this structure when attempting to draw causal inferences, researchers often focus on estimands that are defined relative to a time-varying treatment (e.g., the effect of initiating treatment now versus later). We propose an approach for treatment effect estimation that views indication times as a fixed but possibly unknown pre-treatment covariate (i.e., a variable that is unaffected by assignment to treatment or control) representing relevant characteristics related to the patient's health. 
We then condition on these indication times to construct an estimator of the causal effect of treatment versus control that is independent from the time-structure of the underlying data. 

The paper proceeds as follows. In Section~\ref{sec:background}, we first review existing methods in this domain and describe a general strategy for making causal inferences in longitudinal observational studies by approximating a sequence of randomized trials. 
We then present an alternative conceptualization of the data-generating process that can facilitate causal inference in this setting, which is based on separating the process that governs the time of indication for treatment from the mechanism that determines the receipt of treatment versus control. 
In Section~\ref{sec:framework}, we present a framework for designing longitudinal observational studies to approximate a hypothetical randomized experiment based on this conceptualization and describe the formal assumptions required for inference. 
The core of this framework is a joint state-space model for predicting indication times as a function of longitudinal covariates, described in Section~\ref{sec:model}, which we use to infer the missing indication times for untreated patients. In this Section, we also propose an approach for estimating treatment effects that directly incorporates uncertainty about the inferred indication times. We then illustrate the proposed framework in Section~\ref{sec:app} to study the effects of prescribing contraindicated PDE5Is for treatment of PH using data from the VA health care system.

\section{Background}
\label{sec:background}

When attempting to draw causal inferences in observational studies, it is desirable to ``design" the study in a manner that approximates a randomized experiment \citep{rubin1984william}. 
While this design-based approach to covariate adjustment in observational studies has been widely applied to estimate causal effects using cross-sectional data, little work has focused on extending these methods to longitudinal settings. In fact, how to properly design and analyze longitudinal studies of non-randomized medical interventions (comparative effectiveness studies) remains a point of controversy \citep{rubin2010limitations}. 
A primary concern in this domain is related to the lack of a well-defined control intervention \citep{huitfeldt2015methods}. For instance, in a study evaluating the efficacy of a particular medication, drug A, for treating depression, initiation of control might be marked by the receipt of an alternative medication, drug B. 
Alternatively, if the control intervention is defined as the decision to withhold drug A in favor of a non-pharmacological approach (e.g., self-care or psychotherapy), then receipt of assignment to control may not be available in the observed data. 
In these settings, unknown values of treatment assignment may be missing due to measurement errors (e.g., if receipt of the control intervention is not recorded) or may be (right-)censored due to follow-up (e.g., for patients whose assignment occurs after the study has ended). 
When criteria for determining indications for treatment cannot be applied to the pool of potential controls, an indication time must be inferred for these units.

In many observational studies, controls are selected based on their similarity to treated units with respect to pre-treatment variables observed during a baseline period.
However, control group selection becomes complicated in longitudinal settings where units may enter the study at different points in calendar time. 
Instead of having a single baseline period for all units, in this setting we observe a (potentially unique) baseline period for each treated unit, defined as the time from study entry to the time of treatment initiation. 
The primary challenge is then how to define a comparable pre-treatment period for potential controls who do not receive treatment at any point during the study window \citep{basse2016observational}.
To address this issue, some studies have attempted to identify alternative indication times (i.e., ``phantom" treatment times) for untreated units for whom there is no receipt of treatment.
For instance, in an observational study evaluating the effects of selective cyclo-oxygenase-2 inhibitors for treatment of upper gastrointestinal hemorrhage in eldery patients, \citet{mamdani2002observational} randomly assigned indication times for untreated units.
\citet{zhou2005survival} proposed a similar approach for prescription time-distribution matching, whereby indication times for each untreated unit are selected at random from the distribution of indication times observed among the treated units. 
This strategy ensures that the treatment and control groups will be balanced with respect to time of treatment initiation, but requires the strong assumption that indication times are independent and identically distributed for all units in the study population.

Other studies have addressed this issue by identifying a proxy for the time of treatment indication (e.g., dispensement dates of other prescription medications during the study period)\citep{mcgettigan2006cardiovascular}. 
This approach is the basis of the ``active-comparator design" \citep{yoshida2015active}, which restricts the control group to the set of untreated units who received another active drug during the study period. 
The core idea of this design is that any given sample of untreated units is likely to include subjects with no indications for treatment (e.g., mild disease) as well as subjects with contraindications for the treatment of interest (e.g., due to severe coexisting conditions). 
While this framework provides a useful template for designing comparative effectiveness studies, if the treatment of interest is the most commonly used therapy for a given condition (i.e., a first-line therapy) and the alternatives are used infrequently, selecting an active comparator can be challenging.

Other methods that attempt to address these issues are based on the assumption that, with a sufficient follow-up period, all subjects will receive the treatment of interest \citep{van2011dynamic}.
Under this assumption, an alternative strategy for estimating causal effects is based on framing the data-generating process in terms of a sequence of randomized experiments occurring over time, where at each time the treatment can either be initiated or withheld \citep{robins2000marginal, hernan2008observational}. 
One approach based on this framework is so-called risk set matching (RSM), a technique that was first proposed in \citet{li2001balanced} and has since been widely used in practice (e.g., \citealp{danaei2013observational, kennedy2010effect, watson2019risk}).  
Risk set matching aims to assess the effects of different treatment sequences on an outcome observed at a fixed end-point, thereby avoiding the issue of control group identification. 
Here, the implicit assumption is that, unlike in a classical randomized experiment, all patients will receive treatment eventually and it is rather their \textit{times} of treatment that are a result of randomization. 
Accordingly, causal effects are defined at each of a set of discrete time points and can be estimated by comparing units who received treatment at that time with units who were not yet treated at that point. 
The resulting inferences may be useful to practitioners interested in estimating the effects of delaying treatment with a particular treatment and may help answer questions about the optimal time to apply a medical intervention for patients that require treatment. 
However, the resulting inferences may not be appropriate in settings with treatment by indication, where the time of initiation is not under the control of the treatment initiator, usually the clinician.
When a patient presents with symptoms that indicate the need for medical intervention, their health care provider is faced with a decision about \textit{which} among available treatments to initiate at that time. 
In these settings, it is not sensible to hypothesize about how a patient's outcomes would change if their indications had presented at a different time \citep{angrist1996identification}. 
Rather, the time of indication for treatment should be viewed as a fixed pre-treatment covariate that characterizes the severity of the underlying condition, and causal effects should be framed as contrasts of the outcomes that would be observed for each unit under different treatment strategies initiated at that time.

\section{A causal effects framework}
\label{sec:framework}

The framework we propose mimics the protocol typically used to analyze data from a randomized controlled trial, where outcomes are observed during a pre-specified follow-up period beginning at the time of randomization, which is precisely the same time as initiation of therapy.
To motivate this framework, consider the following hypothetical randomized experiment with a binary treatment, where interest is in estimating the causal effects of treatment on a time-to-event outcome for a particular patient population. 
Here, treatment is defined as the decision to apply the intervention, and the control treatment is defined as the decision to withhold the intervention. 
Assume that patients become eligible for inclusion in the experiment only when their health reaches a point where clinical intervention may be beneficial to the patient despite any associated risks or side effects (the so-called ``indication time"). 
Upon enrollment, suppose that patients are then randomized to receive either treatment or control with probability that depends only on their indication time. 
After randomization, suppose each patient is observed until the earliest occurrence of a pre-specified event (e.g., renal failure) or the end of follow-up, whichever occurs first. The primary outcome is based on this event time calculated with respect to indication time.
In this idealized experiment, contrasts of outcomes between treated and control units with similar indication times will be unbiased estimates of the causal effects of interest. 

This conceptualization of the underlying randomized experiment explicitly defines the control group as the set of untreated patients for whom treatment was indicated during the study period and deliberately withheld (``true controls"). 
The remaining untreated patients are those for whom no indications for treatment were present during the study (``ineligible controls"). 
Because these patients are not assigned to treatment or control during the study period, they are not relevant units for the purposes of causal inference. 
When indication times are fully observed, the set of ineligible controls can easily be identified and discarded prior to analysis. 
However, in many observational studies, indication times may be censored due to follow-up or death and are often only partially observed, specifically, for only those patients receiving treatment. 
In addition, the probabilities of assignment to treatment over time are generally unknown and may be difficult to infer without expert domain knowledge.

\subsection{Notation}

We assume a cohort of $N$ patients (i.e., observational study units), each of whom is observed over a specified study time window divided into $K$ discrete time periods.  
If a patient in the cohort receives the treatment of interest during their observation period, then he or she is included in the study as a treated unit. Other members of the cohort are eligible to be controls during their respective observation periods.

For each unit $i=1,\ldots,N$, suppose we observe a vector of $p$ covariate measurements collected at the $K$ time periods $\boldsymbol{X}_{i} = (X_{i0}, \ldots, X_{iK})$, where $X_{i0}$ is a $p$-vector of baseline covariates observed at study entry. These covariates capture characteristics of each unit or the unit's environment (such as age, gender, physiological factors, diet, medical treatments, and environmental exposures) over the course of the study. Let $T_i$ denote the indication time of unit $i$, which may occur at a discrete time within the study or may be censored at the end of follow-up (i.e., $T_i \in [0,K] \cup [K+1,K+2,\ldots]$). 
By construction, we assume that units whose indication times are censored do not receive either treatment or control and we let $S_i=1\{T_i \in [0,K]\}$ be an indicator of eligibility for inclusion in the study as a control. Similarly, let $M_i$ be an indicator for the missingness of $T_i$ with $M_i=1$ for units whose indication time is unobserved. Finally, let $Z_i$ be an indicator for assignment to treatment upon indication for unit $i$, which equals 1 if unit $i$ is assigned treatment at their indication time $T_i$ and equals 0 if unit $i$ is assigned control upon indication. Unlike the classical setting of a randomized experiment, suppose that $T_i$ is only observed for units who are assigned to treatment at indication times within the study period (i.e., when $Z_i=1$ and $T_i\in [0,K]$).

\begin{table}
	\begin{center}
		\caption{Structure of the observed and missing data. `$\star$' denotes endogenous missingness and `?' denotes exogenous missingness.}
		\label{tab:r0}
		\begin{tabular}{ c | c | c | c | c | c }
			\hline
			Units & $Z_i \in \{0,1\}$ & $T_i \in \{1,2,\ldots\}$ &  $\boldsymbol{X_i}=(X_{i1},\ldots,X_{iT_i})$ & $Y_i(0) - T_i$ & $Y_i(1) - T_i$ \\ \hline
			1 & 1 & $t_1$ & $x_1$ & $\star$ & $y_1-t_1$ \\
			2 & 1 & $t_2$ & $x_2$ &  $\star$ & $y_2 - t_2$ \\
			3 & ? & ? & ? & $y_3-?$ &  $\star$ \\
			4 & ? & $?$ & ? & $y_4-?$ &  $\star$ \\ \hline
		\end{tabular}
	\end{center}
\end{table}

Table~\ref{tab:r0} shows the structure of the data in this setting. Here, it is important to distinguish between two distinct missing data mechanisms that give rise to the observed and missing values. 
The first type of missing data are the indication times that are naturally right-censored at the end of the study. Although, in principle, these values are \textit{observable} over a sufficient follow-up period, the missing-data mechanism is determined by the specified observation period. Because missing data of this type are not ``missing at random" (MAR; \citealp{rubin1976inference}), a model for the missing data mechanism must be incorporated into the analysis in order to yield valid causal inferences \citep{little2019statistical}. The second type of missingness in this setting is due to the fundamental problem of causal inference, which states that we can observe at most one potential outcome  (i.e., the potential outcome corresponding to the treatment actually received) for each subject in the study \citep{rubin1976inference}. These missing potential outcomes are therefore ``endogenous" missing values in the sense that the missingness mechanism is completely determined by treatment assignment. As in a standard randomized experiment, the unit-level missing potential outcomes are impossible to observe under a given treatment assignment, and the goal of causal inference is therefore to recover these values under plausible modeling assumptions in order to make inferences about the causal effects of interest.

\subsection{Causal estimands and assumptions for identification}
\label{subsec:assume}
Interest is in estimating the causal effects of assignment to treatment versus control on time-to-event outcomes defined as the time from indication for treatment to the first occurrence of an event of interest (e.g., death, hospital discharge, symptom remission). 
Under the Rubin Causal Model (RCM; \citealp{holland1986statistics}), each participant has two potential outcomes, $Y_{i,T_i}(0)$ and $Y_{i,T_i}(1)$, which represent the outcomes that would be observed for unit $i$ under assignment to control or treatment, respectively, where assignment occurs at exactly the indication time $T_i$.  
Let $Y_T(0)=(Y_{1T}(0),\ldots,Y_{NT}(0))$ and $Y_T(1)=(Y_{1T}(1),\ldots,Y_{NT}(1))$, and let $Y_T=(Y_T(0),Y_T(1))$ denote the complete set of potential outcomes for all units relative to indication time $T$. 
We make the Stable Unit Treatment Value Assumption (SUTVA, \citealp{rubin1980randomization}), which asserts that there is no interference between units and no hidden forms of treatment. 
Under this assumption, the average treatment effect (ATE) at a single indication time $T$ is defined as:
\[\tau_T = \EE{Y_{T}(1) - Y_{T}(0)},\]
where $\EE{\cdot}$ is the average across all units. 
In longitudinal studies where indication may occur over a fixed study period $[0,K]$, we can construct an aggregate average measure of these time-specific effects as 
\[\tau = \frac{1}{K}\sum_{t=1}^K \tau_t. \]	

In settings where the outcome of interest $Y$ is defined relative to a time of death or failure, $\tau$ captures the average change in survival time under treatment compared to control for units who present with indications for treatment at similar times over the study period. 
Alternatively, as we will see in Section~\ref{sec:app}, causal estimands can also be specified to quantify the average difference in survival rates at particular time points during the follow-up period. 
To construct an unbiased estimator of $\tau$ using non-randomized data, we assume that treatment assignment is conditionally independent of the potential outcomes given all pre-treatment covariates including the indication times (i.e., $(Y_{i,T_i}(0), Y_{i,T_i}(1)) \independent Z_i | X_{iT_i}, T_i$). 
In the clinical context we consider, this asserts that assignment to treatment versus control is unconfounded given indication times, $T$, and observed pre-treatment covariates, $X$. 
Under this assumption, an unbiased estimate of the treatment effect is:
\begin{align}
\label{eq1}
\hat{\tau}
&=\frac{1}{N_1}\sum_{i:Z_i=1} Y_{iT_i}(1) -  \frac{1}{N_0}\sum_{i:Z_i=0} Y_{iT_i}(0),
\end{align}
where $N_1=\sum_{i=1}^N S_iZ_i$ is the total number of treated units in the cohort and $N_0 = \sum_{i=1}^N S_i(1-Z_i)$ is the total number of true controls.
Here, in contrast to the classical setting, the outcomes that are actually observed for each unit depend not only on their treatment assignment but also on their indication time.
For units whose indication times are not observed, these missing times must be inferred in order to evaluate \eqref{eq1}..
Note that with an untreated comparison group, estimating the average treatment effect for the treated (ATT) using techniques such as propensity score matching may be preferred, since we would rarely want to initiate treatment for all units in the population.
However, in the setting of treatment by indication or when an active comparator is present (e.g., comparative effectiveness research), estimating the average treatment effect for the entire population is desirable because we are focusing on the choice of one treatment compared to another assuming that indications for treatment have already occurred. 

\subsection{Overview of inferential approach}

Our conceptual framework implies by design that the missingness of the times of indication, $T^{mis}$, and the indicators for assignment to control, $Z=0$, are completely dependent on the values of those missing measurements. 
That is, we assume that the data $(T_i,Z_i)$ for unit $i$ will be missing if $T_i>K$ (regardless of the value of $Z_i$) or if $Z_i=0$ (regardless of the value of $T_i$). This means that the missing measurements (indicated by $M_i=1$) are ``missing not at random" (MNAR; \citealp{imbens2015causal}). 
By inferring the missing indication times for the untreated units, we can minimize the information loss that arises from the missing data mechanism to make more precise inferences from the observed data.

The first goal is therefore to construct a model for predicting the observed indication times $T^{obs}$ based on baseline and time-varying covariates $X$, which we will use to infer the missing indication times. This model will also induce a probability distribution on the potential outcomes in the control group, since the observed outcomes must be calculated relative to the indication times. By applying the assumptions described above in Section \ref{subsec:assume}, we can separate the joint distribution of the complete data, including all observed potential outcomes $Y_T^{obs}$ as well as both the observed and unobserved indication times, $T=(T^{obs},T^{mis})$ given some global parameter $\theta$ partitioned as $\theta=(\theta_1,\theta_2,\theta_3)$ as
\begin{equation}
 p(Y_T^{obs},T,Z|X,\theta) = p(Y_T^{obs}|T,Z,X,\theta_1) p(Z|T,X,\theta_2)p(T|X,\theta_3).
 \end{equation}
For Bayesian inference with prior density $p(\theta)=p(\theta_1,\theta_2,\theta_3)$, the posterior density of $\theta$ given the complete data is given by 
\begin{equation}
p(\theta|Y_T^{obs},T,Z) \propto p(Y_T^{obs}|T,Z,X,\theta_1) p(Z|T,X,\theta_2)p(T|X,\theta_3)p(\theta_1,\theta_2,\theta_3).
\end{equation}
Posterior inference on $\theta$ can then proceed by straightforward application of Markov Chain Monte Carlo (MCMC) techniques, such as the Gibbs sampler \citep{geman1984stochastic, gelman2014bayesian}. For example, in each iteration of the Gibbs sampler, we draw the missing indication times $T^{mis}$ from the conditional posterior predictive distribution of $T^{obs}$ given covariates $X$ and the current draw of the parameter $\theta$. 

The completed indication times can then be used to classify untreated patients into distinct groups of true controls and ineligible controls based on eligibility $S$, where the true control group consists of patients with $M_i=1$ and $S_i=1$. For the true controls, we can then calculate values for the potential outcomes $Y_T(0)$ given the generated values of $T$. These values are then regarded as observed potential outcomes, denoted $Y_T^{obs}$, which are equal to the calculated $Y_T(0)$ for units classified as true controls and equal to $Y_T(1)$ for treated units. Given the observed potential outcomes $Y_T^{obs}$, the completed times of indication $T=(T^{obs},T^{mis})$ and the corresponding assignment vector $Z$, we can then update the parameters $\theta_1,\theta_2$ and $\theta_3$ by drawing from the conditional posterior distribution,
\begin{equation}
p(\theta_1,\theta_2,\theta_3|Y_T^{obs},T,X,Z) \propto p(Y_T^{obs}|T,Z,X,\theta_1)p(Z|T,X,\theta_2)p(T|X,\theta_3)p(\theta_1,\theta_2,\theta_3).
\end{equation}

Posterior inference on the causal effects of interest can be obtained by computing the values of the constructed estimator within each MCMC iteration and summarizing their distribution across the posterior sample. 
Thus, in each iteration, we can construct a dataset consisting of the observed indication times, the simulated indication times, and all observed potential outcomes, and use these completed data to calculate an estimate of the treatment effect as in Equation~\eqref{eq1}. 
Alternatively, we could specify a joint distribution for the potential outcomes $Y_T=(Y_T(0),Y_T(1))$ that we could then use to impute the missing potential outcomes $Y_T^{mis}$ in each iteration by drawing from the conditional distribution with density function $p(Y^{mis}_T|Y^{obs}_T, T, X, \theta)$. Repeating this process over many such simulated datasets produces the approximate posterior distribution for all causal effects of interest. In the same way, posterior samples of $\theta$ can provide posterior estimates of the parameters that characterize the data-generating process; this is described in greater detail in Section~\ref{sec:model}.

\section{State-space model for time of treatment indication}
\label{sec:model}

Based on the conceptual framework presented in Section~\ref{sec:framework}, we propose a specific and pragmatic model for predicting the time of indication for treatment - the earliest time at which a patient presents with indications for clinical intervention - as a function of both fixed and time-varying covariates. In particular, we hypothesize that observed covariate measurements that reflect worsening health and diminished functional capacity will be predictive of indication times. Similarly, we assume that covariates capturing provider characteristics or temporal features (e.g., month or year when measurements are recorded) are independent of the indication times but may influence the probability of assignment to treatment upon indication. For instance, in the application presented in Section~\ref{sec:app}, we expect the probability of treatment to decline over time as clinicians become more informed about populations where the medication of interest may be contraindicated. We capture these separate dependencies through separate components of a hierarchical model; the first component characterizes the stochastic process that governs patients' indication times, and the second component describes the conditional probability of assignment to treatment given the indication times. Specifically, we model a patient-level health process generated by time-varying covariates together with random fluctuations over time and adopt a so-called ``threshold approach" \citep{albert1993bayesian}, which views the indication time as the first hitting time of this latent process. Similarly, we assume that the conditional probability of assignment to treatment versus control given the indication time for each patient varies based on institutional preferences, which may systematically change over time with widespread changes in the established guidelines for treatment.

\subsection{Model formulation}

Suppose that the time series of covariate measurements for each unit, $\boldsymbol{X}_{i,1:K}$, is independent from the measurements of other units, and let $\theta_{i,1:K}=(\theta_{i1},\ldots,\theta_{iK})$ be a state variable representing fluctuations in unit $i$'s overall health over the course of the study that are not explained by the covariates. We assume a normal distribution for the daily health fluctuations for unit $i$ between time periods $t$ and $t-1$, such that
\begin{equation} 
\label{eq:theta}
\theta_{it}=\rho \theta_{i,t-1}+\epsilon_{it}, \hspace{3mm} \epsilon_{it}\sim \mathcal{N}(0,\sigma_{\epsilon}^2),
\end{equation}
where $\mathcal{N}$ denotes the normal distribution. 
To make the model identifiable in our setting with a binary observation process, we fix $\sigma_{\epsilon}^2=1$.
We assume $|\rho|<1$ (i.e., the latent state process $\theta_{1:T}$ follows a stationary autoregressive model of order one).
Thus, the model balances short-term changes in health status with information from long-term health trajectories.

To initialize this process, we assume a standard normal prior distribution for $\theta_{i0}$. 
Here, the stochastic component $\epsilon_{it}$ captures the unexplained variation of unit $i$'s health over time.
Conditional on the latent states $\theta_{i,1:K}$, we then define the observation process $\Psi_{i,1:K}$, which relates the overall health trajectory of unit $i$ to their indication time according to a probit regression model as
\begin{equation}
\label{eq:psi}
P(\Psi_{it}=1|\theta_{it},X_{it})=\Phi(\theta_{it} + X_{it}\beta),
\end{equation}
where $\Phi(\cdot)$ denotes the cumulative distribution function of a standard normal random variable and $\beta \in \mathbb{R}^p$ is a vector of regression coefficients.
Under the latent variable representation of \eqref{eq:theta} and \eqref{eq:psi} (also referred to as a state-space mixed model; \citealp{czado2008state}), the indication time for each unit $i$ can be expressed as 
\begin{equation}
\label{eq:time}
 T_i= \inf \{t \in [0,K]: \Psi_{it}=1\}.
 \end{equation}

Thus, the model for indication times $T_i$ induces a conditional distribution for the time of indication given a vector of longitudinal, pre-treatment covariates. One of the main advantages of this latent state representation is that it can flexibly accommodate both fixed and time-varying covariates. The proposed modeling approach can also be extended to settings with non-linear covariate effects by replacing the linear effects in \eqref{eq:psi} with nonlinear contributions, such as those described in \citet{denison2002bayesian}.

We consider a separate model for the assignment mechanism, which determines assignment to treatment versus control upon indication. Here, our model formulation is based on the assumption that, in many settings, variation in treatment practices may be due to clinician and/or institutional preferences rather than differences in patient characteristics \citep{slaughter2017comparative}. In particular, we assume that each unit is assigned to treatment with a probability that depends on their indication time. 
Conditional on $T_i$, and on exogenous covariates $D_i$, 
the assignment mechanism  for our specific model can be expressed as
\begin{equation}
\label{eq:prob}
 Z_i|T_i,D_i \sim Bernoulli(\pi_{iT}), \hspace{4mm} \text{logit}(\pi_{iT})=
 \delta_0 + f(T_i, D_i, \delta_1),
 \end{equation}
where  $f(\cdot)$ is a parameterized, possibly non-linear, function.
deterministic transformation of the indication time (e.g., calendar
year or month). 
This model can be easily extended to accommodate 
other artifacts of the study that are believed to influence the probability of treatment assignment. 
For example, in Section~\ref{sec:app}, we 
let $ \text{logit}(\pi_{iT})= \delta_0 + \delta_1 D_i$
where $D_i$ is calendar time, and 
$\delta_1$ is restricted to be negative so that the probability of 
receiving the treatment is monotonically decreasing over time. 

\subsection{Inferential procedure}

Our procedure uses the Kalman Filter  \citep{carlin1992monte} to marginalize out the latent state parameters $\theta_{1:T}$ for more efficient inferences. The full likelihood of the parameters $\Omega=(\Psi_{1:K},\rho,\beta,\delta_0, \delta_1)$ can be written as 
\begin{equation}
\label{eq:llik}
\mathcal{L}(\Omega) \propto \prod_{i=1}^N \left( p(T_{i}^{obs}=t|\Omega)p(Z_{i}=1|T_{i}^{obs}=t,\Omega)\right)^{1-M_i} \left(p(M_i=0|\Omega)\right)^{M_i}
\end{equation}
where
\[p(M_i=0|\Omega) = 1 - \sum_{t=0}^K p(T_{i}^{mis}=t|\Omega)p(Z_{i}=1|T_i^{mis}=t,\Omega).\]
Letting
$\boldsymbol{Z}$ and
$\boldsymbol{M}$
be the vectors of the $Z_i$ and $M_i$ across the $N$ patients,
the associated posterior density is therefore
\begin{align*}
p(\Omega|T^{obs},\boldsymbol{Z},\boldsymbol{X},\boldsymbol{M}) 
&\propto 
p(\Omega) \prod_{i=1}^N
p(T_i^{obs}|\boldsymbol{X},\Omega)p(Z_i|T_i^{obs},\Omega)p(M_i=0|\Omega).
\end{align*}
Our modeling framework permits a flexible choice of prior distributions.
For example, 
we can assume a prior distribution that factors into independent densities as
\begin{align*}
p(\Omega) &= p(\rho,\beta,\delta_1,\delta_0, \Psi_{1:K})\\
&= p(\rho)p(\beta)p(\delta_1)p(\delta_0)p(\Psi_{1:K}|\rho,\beta).
\end{align*}
Many distributional choices are possible for each prior component.
One flexible set of choices might be
\begin{align*}
p(\rho) &\sim \mathcal{N}(\rho_0,\sigma_{\rho}) \hspace*{0.2in} 
\mbox{for $-1 \leq \rho \leq 1$} \\
p(\beta) &\sim \mathcal{N}_{p}(\beta_0, \Sigma_0) \\
p(\delta_0) &\sim \mathcal{N}(a,b)\\
p(\delta_1) &\sim Gamma(c,d) \hspace*{0.2in} \mbox{if $\delta_1$ is restricted to be positive}\\
p(\Psi_{1:K}|\rho,\beta)&= p(\Psi_{1}|\rho,\beta)\prod_{j=2}^K p(\Psi_{j}|\Psi_{j-1},\rho,\beta)
\end{align*}

Posterior sampling is
straightforward to implement using standard software for implementing MCMC methods such as JAGS \citep{plummer2003jags}. 
Rather than sampling each $\Psi_t$, $t=1,\ldots,K$, conditional on the rest, it is more efficient
to sample them jointly via MCMC using the Kalman filter \citep{carlin1992monte}.
As previously described, these posterior samples allow us to measure the missing times of treatment assignment and assignment to control according to \eqref{eq:time}, where $T^{mis}>K$ for units with $\Psi_{1:T}=0$. Given the inferred indication times, our framework implies that $Z_i=0$ for units with $T_i^{mis} \in [0,K]$. Recall that treatment assignment is undefined for those whose time of treatment assignment was censored by the end of the study (i.e. units with $T_i^{mis}>K$). Finally, we can calculate the observed outcomes for units inferred to belong to the group of true controls by computing the difference between the observed event times and the inferred indication times for those units.

\subsection{Bayesian analysis with inferred indication times}
One of the benefits of the proposed approach is that it allows us to make inferences about the missing indication times that are free from the outcome analysis. Our approach allows for flexible specification of the causal estimands of interest and also allows researchers to choose any mode of inference for analysis of the outcomes that they see appropriate. 
For example, one might use the posterior mode of inferred times of indication for each unexposed unit calculated over a large number of MCMC samples as the point estimate for that unit's indication time. 
This can then be viewed as a single imputation of the missing values, and conditional on these estimates one could estimate the treatment effects by simple comparisons of means of the time-to-event outcomes (e.g., using a Neymanian or Fisherian mode of inference).
Alternatively, our framework also allows for more sophisticated analysis of outcomes. For instance, we could first obtain a large number of posterior samples for the missing times of indication across for all untreated units and use these samples to form unit-level empirical posterior distributions of the missing indication times. Then, one might iteratively impute values from these distributions and calculate the values of the corresponding observed potential outcomes in each iteration. 
Contrasts between treatment and control groups evaluated across all iterations would then produce a distribution for the causal estimate (given the potential outcomes) that incorporates uncertainty about the missing indication times.

It is important to note is that this type of approach is only partially Bayesian in that we are using Bayesian methods to infer missing values that are needed in order to measure the observed outcomes. 
To make this a fully Bayesian approach, as previously mentioned, one could also specify a model for the missing potential outcomes (i.e., the missing time-to-event outcomes under alternative assignment) conditional on the partially observed outcomes, the assignment vector, and/or the covariates. 

\section{Application}
\label{sec:app}

To illustrate our proposed methods, we analyzed data from a recent study using electronic medical records from the VA health system evaluating the impacts of inappropriate prescribing practices for the treatment of pulmonary hypertension \citep{kim2018phosphodiesterase}. 
Pulmonary hypertension is a condition of high blood pressure that affects arteries in the lungs and heart. One common treatment for PH is a class of medications called phosphodiesterase-5-inhibitors (PDE5Is), which act on enzymes causing blood vessels to relax in order to lower blood pressure. While PDE5Is have been shown to be effective for treating some rare forms of PH, \citet{freiman2015patterns} identified the use of these drugs as wasteful, ineffective, and potentially harmful for treating patients with more common types of PH caused by left heart disease (WHO Group 2) or hypoxemic lung disease (WHO Group 3). Despite its contraindication, a study of veterans diagnosed with these types of PH over the years of 2005 to 2012 identified over 2,000 prescriptions for PDE5Is that were inappropriately administered to patients in the VA health system. To understand the impact of these inappropriate treatment practices on patient outcomes, it is of interest to measure the causal effects of prescribing PDE5Is to patients with PH Groups 2 and 3 on the time lag between the application of the intervention and the occurrence of a clinical event of interest (e.g., time from treatment to death).

The primary question that we sought to address through this analysis concerned the extent to which treatment with PDE5Is for patients diagnosed with PH Groups 2 and 3 impacted the one-year survival rate compared to patients who had similar indications for treatment but were not prescribed PDE5Is during the study period. As a secondary objective, we were also interested in inferring which among a large set of observed patient and provider characteristics were strongly associated with an increased risk of receiving a contraindicated PDE5I for treatment of PH. 

\subsection{Data}
Our data set contains demographic and laboratory measurements as well as records of the utilization of medications, inpatient and outpatient services for over 350,000 veterans who were diagnosed with PH Groups 2 and 3 and received prescription medications from the VA from 2005 to 2016. For all patients, baseline health-related measurements were collected at the 
time of first PH diagnosis based on ICD-9 diagnosis codes within either the VA health system or Medicare. Subsequent health-related measurements were collected at intermittent observation times corresponding to patient-provider clinical interactions (e.g., an inpatient or outpatient visit). The exact number of measurements recorded and the time elapsed between subsequent measurements varied by patient. Observations with implausible values for lab measurements or demographic variables were excluded prior to analysis.

For the present analysis, we considered patients who at the time of PH diagnosis (Group 2 or 3) were between 65 to 95 years of age, were eligible for Medicare benefits, and who had not received prescriptions for a PDE5I medication prior to PH diagnosis. 
Because female patients comprised less than 3\% of the patient population who met these initial eligibility criteria, we further limited our analysis to only male patients. 
For data integrity purposes, we also excluded any patients who were receiving Medicare part C at the time of diagnosis as well as patients who did not fill any prescriptions within the VA health system during the one year prior to their diagnosis, since these patients may have received PH related care from private providers that we could not track. 
Finally, we excluded any patients who received a prescription for nitrates after PH diagnosis and prior to a PDE5I prescription because nitrates are contraindicated for treatment with PDE5Is. 
Among patients who met all initial eligibility criteria, we defined the ``treatment group" as the set of patients who filled at least one prescription during the study period for a PDE5I medication primarily indicated for the treatment of PH. 
In particular, we excluded medications with secondary or off-label indications for the treatment of PH such as sildenafil, tadalafil, and verdenafil.

Because we observed the prescription dispense dates rather than the dates when the medications were first prescribed, we excluded from our analyses any treated patient whose first PDE5I prescription was dispensed more than 60 days after a hospital visit. 
After applying all exclusion restrictions, the remaining sample was comprised of 534 patients who received treatment for PH with a PDE5I medication within a one-year period following their diagnosis date and 167,701 potential controls who did not receive a PDE5I during that period. 
Our outcome of interest is survival time in the period following indication for treatment, which we observed for treated units and must be inferred for units in the control group. We consider the date of PH diagnosis as the time of ``earliest eligibility" for indication. For each potential control patient, we base our inferences on clinical data observed at intermittent intervals from the time of earliest eligibility until death or the end of follow-up in December 2016, whichever occurred first.

\subsection{Constructing comparison groups and time-varying covariates}
To make the assumptions of our proposed framework described in Section~\ref{sec:framework} more plausible in this setting, we selected a set of potential control units who appeared similar to the treatment group at baseline based on health-related measurements collected in the 1 year prior to PH diagnosis (excluding the date of diagnosis). We considered a set of $17$ baseline covariates identified as predictive of the time of assignment to treatment within the treatment group. See Appendix~\ref{app:covs} for details of this variable selection procedure. Potential controls were selected using 1:1 nearest-neighbor matching with replacement \citep{rubin1973matching}, whereby each treated unit is matched to its closest control unit based on the Mahalanobis distance calculated over baseline covariates.

This produced a final sample of 534 treated units and 531 matched control units who were similar to the treatment group at their times of PH diagnosis but did not receive a PDE5I at any point during the observation period. Diagnostics performed after matching confirmed that the covariate distributions were adequately balanced between the treatment group and the matched potential control group. Table~\ref{tab:covs1} shows descriptive statistics on baseline variable for the final matched samples. After matching, we proceeded under the assumption that overall health status was unconfounded given baseline covariates such that patients with similar covariates at baseline were expected to have similar health trajectories and therefore, similar indication times over the course of study.

\begin{table}[htbp]
	\centering
	\caption{Baseline covariate data for potential controls and treatment group.}
	\label{tab:covs1}
	\begin{tabular}{lcc}
		\hline
		& \multicolumn{1}{l}{Potential Control} & \multicolumn{1}{l}{Treatment} \\
		\hline
		PH Group 2 & 88\%  & 88\% \\
		Recently Hospitalized & 13\%  & 12\% \\
		Recent Procedure & 65\%  & 75\% \\
		Recent ER Visit & 52\%  & 53\% \\
		Age   & 74.8  & 74.5 \\
		Weight & 203.3 & 204.5 \\
		Height & 69.5  & 69.5 \\
		Resting Heart Rate & 74.2  & 75.1 \\
		Systolic Blood Pressure & 130.1 & 130.5 \\
		Diastolic Blood Pressure & 71.2  & 71.2 \\
		Inpatient Days  & 1.19  & 1.22 \\
		Outpatient Days & 27.3  & 27.9 \\
		Number of Comorbidities & 0.52  & 0.54 \\
		Cardiac Events & 1.60  & 1.49 \\
		Pulmonary Events & 0.61  & 0.69 \\
		Organ Failure Events & 1.09  & 1.16 \\
		Number of Medications & 11.4  & 11.8 \\\hline
	\end{tabular}%
\end{table}%

In addition to baseline covariates, we also included in our analyses a number of time-varying covariates collected at intermittent observation times throughout the study period. 
Specifically, we considered six time-varying, clinically meaningful covariates that indicate changes in patients' disease severity and health status over time, including an indicator for whether the patient was most recently observed in an outpatient versus inpatient setting, an indicator for the presence of new comorbidities, and separate indicators for recent hospitalization, organ failure events, cardiac events or pulmonary procedures recorded during the 30 days prior to each visit. 
We also included as a single time-varying covariate the Mahalanobis distance calculated between each patient's laboratory measurements (e.g., heart rate and blood pressure) at baseline and values of the same laboratory variables collected at each follow-up visit.
This allowed us to greatly reduce the dimension of the covariate space and operationalize the laboratory variables in terms of gain scores. 

We observed a strong positive correlation between the empirical probabilities of survival and indication times for patients in the treatment group. Among 77 patients who received treatment within 7 days following their PH diagnosis (i.e., patients with $T_i \leq 7$), only 53 patients (68.8\%) were alive one year post-diagnosis. Averaged over all 534 patients whose indication times occurred within one year of diagnosis (i.e., the treatment group), this survival rate increased to 80.6\%. In contrast, the survival rate at one-year after PH diagnosis for the 531 matched potential controls was approximately 81.7\%; however, because we do not observe the indication times for these patients, this is a naive estimate that is likely negatively biased. Also note that these survival probabilities are defined relative to \textit{diagnosis times}, which we regard as a fixed pre-treatment covariate. As a result, these measurements are purely descriptive and should not be regarded as treatment effects.

\subsection{Results}

Using the approach described in Section~\ref{sec:framework}, we fit the model defined by equations \eqref{eq:theta}--\eqref{eq:prob} using MCMC posterior simulation. 
We constrained $\delta_1$, the effect of calendar time on the log-odds of PDE5I initiation, 
to be negative for this application since the probability of receiving PDE5Is was believed to be monotonically decreasing over the course of the study. This is because PDE5Is were believed to be the standard, first-line therapy for treatment of PH at the beginning of the study with use steadily decreasing as knowledge of its contraindication for some PH patients spread throughout the medical community.

To estimate the treatment effects of interest, we fit eight distinct models.
Each model assumed differing time windows within which the indication time might occur, or that (for some untreated units) the indication time occurred beyond the time window.
Specifically, we assumed time windows corresponding to indication times occurring within the first 14 days, 30 days, 60 days, 90 days, 120 days, 180 days, 270 days, and 365 days (1 year) after PH diagnosis. 
Our health process model component assumed that changes could occur daily, so that the number of time periods
$K$ varied from $14$ to $365$.
For each interval, we estimated the average causal effects of treatment versus control on a binary measure of survival at follow-up one year after indication. In addition, we estimated the conditional average treatment effect (CATE) for each of the eight time intervals, where in each interval the average is over the subset of patients whose indication times were within a specified range (e.g., the conditional average treatment effect for patients whose indication times occurred within 15-30 days following PH diagnosis). 
To estimate these effects, we first identified the subset of units whose indication times (either observed or inferred) were within the specified study period based on the current values sampled using MCMC. Within this subset, we then calculated survival rates at one year post-indication for treated units as well as for units classified as true controls. Finally, we measured the treatment effect as the difference between these two proportions. For each of the specified study periods, we ran the MCMC sampler with four parallel chains each run for 20,000 iterations, where the first 5,000 draws of each chain were discarded as a burn-in period. With the resulting 60,000 samples, we calculated the posterior means and 95\% credible intervals (CI) for all model parameters. In all cases, the MCMC simulated model parameters and quantities of interest 
raised no concerns using standard diagnostics including those by
\citet{geweke1992evaluating} and \citet{gelman1992inference}. As an additional sensitivity check, we evaluated the performance of the proposed model in each setting under different choices of hyperparameters using the deviance information criterion (DIC; \citealp{spiegelhalter2002bayesian}) and found the results to be generally unaffected.

\begin{figure}[ht]
	\begin{center}
		\includegraphics[width=0.75\textwidth]{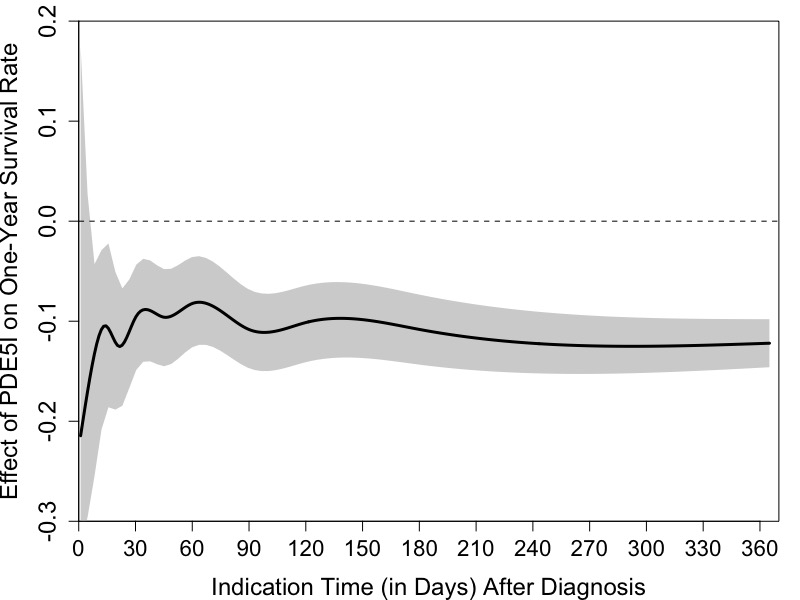}
	\end{center}
	\caption[Estimated effects of PDE5I on one-year survival]{Cubing smoothing spline curve for estimated effects of PDE5I compared to control on one-year survival rate based on time from diagnosis to indication. Shaded area shows pointwise 95\% posterior credible interval.} 
	\label{fig:app2}
\end{figure}

Posterior means and 95\% credible intervals for the treatment effects of interest are presented in Table~\ref{tab:results1}. Here, the causal estimand at each time point is defined as the average effect of treatment on one-year survival for all patients with indication times occurring at or before that time (i.e., $\hat{\tau}_K = E[Y_i(1)-Y_i(0)|T_i \leq K]$ for $K=14, 30, 60, 90, 120, 180, 270, 365$). Table~\ref{tab:results1} also summarizes the cumulative number of potential control units identified as ``true controls" at each time point based on the posterior median of the 60,000 MCMC samples (i.e., $\hat{N}_{0K}=\sum_{Z_i=0} 1\{\hat{T}_i \leq K\}$). 
Figure~\ref{fig:app2} shows the smoothed effects of treatment over a one year period following PH diagnosis, calculated using a 
cubic smoothing spline \citep{marsh2001spline}, which we specified with eight knots located at each posterior mean treatment effect estimated using our model.

\begin{table}[t]
	\centering
	\caption[Estimated effects of PDE5Is based on times of indication]{Posterior median number of true control units inferred from matched sample of potential controls at each time point and estimated effects of treatment compared to control with 95\% posterior intervals.}
\label{tab:results1}
	\begin{tabular}{l|ccc|cc|cc}
		\hline
		\multicolumn{1}{l|}{Study}& Number &  \multicolumn{2}{c|}{Inferred number} & \multicolumn{2}{c|}{One-year survival} &\multicolumn{2}{c}{Estimated impact}\\
		\multicolumn{1}{l|}{period}& treated &  \multicolumn{2}{c|}{of ``true controls"} & \multicolumn{2}{c|}{after indication} &\multicolumn{2}{c}{of PDE5I} \\ \cline{3-4} \cline{5-6}\cline{7-8}
		\multicolumn{1}{l|}{($K$)}& ($N_1$) & $\hat{N}_{0K}$ & 95\% CI & Treatment& Control& $\hat{\tau}_K$ & 95\% CI \\ \hline
		14 Days & 109& 48 & (34, 62) &71.6\% & 82.0\% & -10.4\% & (-18.6\%, -1.6\%) \\
		30 Days & 164 & 88  & (72, 103) & 73.2\%& 82.8\% & -9.6\% &  (-14.5\%, -3.8\%)\\
		60 Days & 233 & 113 & (96, 127) & 75.1\% & 83.3\% & -8.2\% & (-12.8\%, -5.0\%) \\
		90 Days & 280 &  127 & (111, 141)& 72.5\%& 83.3\% & -10.8\% & (-14.7\%, -6.8\%) \\
		120 Days & 318 & 134 & (119, 147) & 73.0\% & 85.1\% & -10.1\% & (-14.1\%, -6.4\%)\\
		180 Days & 380 & 150 & (135, 162) & 73.6\%& 84.3\% & -11.0\% & (-14.1\%, -7.5\%) \\
		270 Days & 456 & 165 & (151, 180)& 74.0\%& 84.5\%& -10.5\% & (-13.9\%, -7.5\%)\\ 
		365 Days & 534 & 183 & (170, 195) & 72.5\% & 84.5\% & -12.0\% & (-14.6\%, -9.8\%)  \\ \hline
	\end{tabular}%
\end{table}%
\clearpage

These results suggest that the majority of the matched potential controls were ineligible to receive treatment during the one year period following their PH diagnosis. For these patients, lack of treatment can therefore be interpreted as a lack of \textit{indication} for treatment. On the other hand, potential controls with inferred indication times occurring at or before each specified time point are regarded as ``true controls" for whom, upon indication for treatment, PDE5Is were actively withheld, possibly in favor of alternative medication or treatment strategy. Given the inferred times of indication, these patients provide a credible comparison group with which we can make causal inferences. In particular, our findings indicate that among patients with PH (Groups 2 and 3) who are indicated for treatment at any point in the one year following their diagnosis, treatment with PDE5Is has a large, negative effect on one-year survival probability.

Table~\ref{tab:results2} shows posterior estimates for the baseline and time-varying covariate effects, $\beta_1$ and $\beta_2$, as well as for the correlation between patients' latent health states, $\rho$, and parameters governing the probability of assignment to treatment upon indication, $\delta_0$ and $\delta_1$. Here again, our results are largely consistent across the settings with credible intervals shrinking as study period widens. 
In general we find a small negative posterior correlation between latent health measurements, which decreases in magnitude over time. This suggests there may be some variation in patients' overall health that is not explained by the observed covariates in the time period immediately following PH diagnosis, but this systematic variation dissipates over time.

\begin{figure}[!htbp]
	\begin{center}
		\includegraphics[width=0.75\textwidth]{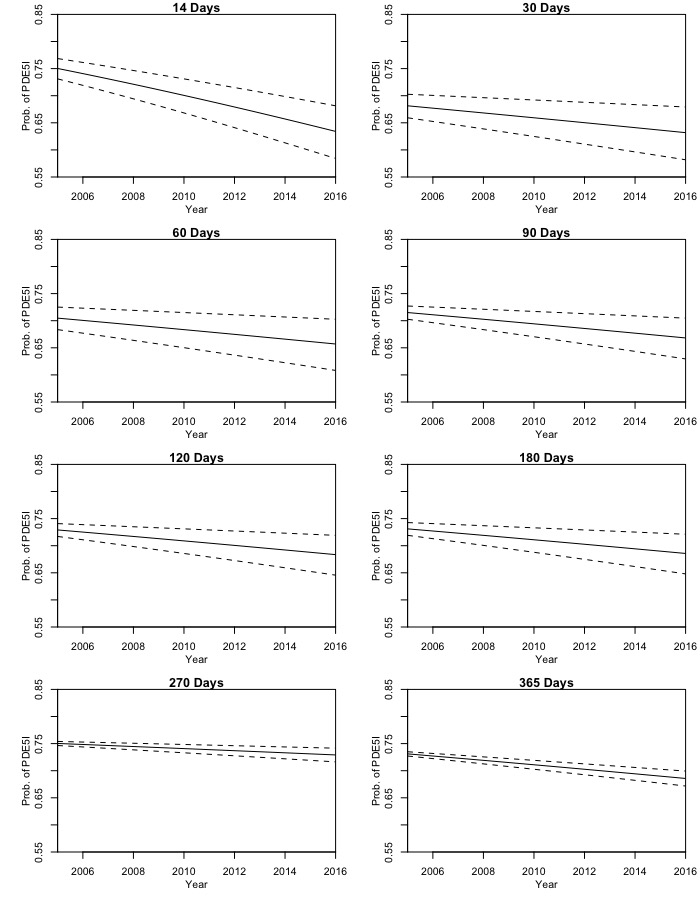}
	\end{center}
	\caption{Cubic smoothing spline curves showing the conditional probability of receiving PDE5I based on time of indication for eligible patients in the VA health care system diagnosed with PH between 2002 and 2016. Dashed lines show 95\% posterior credible intervals.}
	\label{fig:probtreat}
\end{figure}

Posterior estimates for parameters of the treatment assignment mechanism are also relatively stable across the different study periods. Baseline probabilities of treatment upon indication range from $0.79$ for the study period of 14 days to $0.73$ for the study period of 365 days, with the probability of treatment slowly decreasing over time.
Figure~\ref{fig:probtreat} shows this trend over time for eligible patients in the VA health care system diagnosed with PH between 2002 and 2016 based on posterior median estimates of $\delta_0$ and $\delta_1$ within each study period.

A key feature of our model is that it allows us to directly evaluate which of the baseline and time-varying covariates carry more or less information about patients' times of indication. In the present study, posterior inferences for covariate effects within each of the study periods were similar. In general, our findings suggest that patients with PH Group 2 were more likely to have indications for treatment shortly after PH diagnosis than patients with PH Group 3. Results also indicate that occurrence of one or more incidental medical procedures (e.g., cardiac surgery or pulmonary function testing) within 30 days prior to PH diagnosis is strongly associated with earlier indication times. Further, we found that patients who regularly receive care in an inpatient setting generally have earlier indication times than patients whose follow-up visit typically occur in an outpatient setting. 
Among the other baseline covariates included in our analysis, occurrence of one or more pulmonary disease events (e.g., pneumonia) within 30 days prior to PH diagnosis and number of comorbidities present at baseline were also positively associated with indication for treatment during the study period. These results may offer insights for clinicians about best practices for health management of PH patients, and may also be used to guide modeling decisions in other applications.

\begin{landscape}
\begin{table}[htbp]
  \centering
	\caption[Estimates of model parameters based on MCMC]{Posterior medians with 95\% credible intervals for parameters $\rho$, $\delta_0$, $\delta_1$, $\beta_1$ (baseline covariate effects) and $\beta_2$ (time-varying covariate effects) in each study period.}  
	\label{tab:results2}  \begin{tabular}{llcccccccc}
\hline
    \multicolumn{2}{l}{Parameter} &\multicolumn{1}{c}{14 Days} & \multicolumn{1}{c}{30 Days}  &\multicolumn{1}{c}{60 Days} & \multicolumn{1}{c}{90 Days}  &\multicolumn{1}{c}{120 Days} & \multicolumn{1}{c}{180 Days}  &\multicolumn{1}{c}{270 Days} & \multicolumn{1}{c}{365 Days} \\ \hline
    \multicolumn{10}{l}{$\beta_1$ (Baseline covariates)}  \\ 
        & PH Type 2\footnotemark[1] & 1.588  & 1.759  & 1.851  & 1.901  & 1.958  & 1.961  &  2.04     & 2.18 \\
    &History of organ failure\footnotemark[1]\footnotemark[2] & 0.196  & 0.215  & 0.236  & 0.259  & 0.278  & 0.275  &    0.294   &  0.315 \\
    &Recently hospitalized\footnotemark[1]\footnotemark[2] & 0.042  & 0.017  & -0.016 & -0.084 & -0.122 & -0.126 &  -0.165     & -0.210 \\
    &Age  & -0.103 & -0.079 & -0.086 & -0.095 & -0.085 & -0.087 &  -0.071     & -0.050  \\
    &Comorbidities & 0.253  & 0.256  & 0.369  & 0.406 & 0.445  & 0.441  &  0.398     & 0.430 \\
    &No. of recent operations\footnotemark[2] & 0.779  & 0.911  & 1.023  & 1.072  & 1.120  & 1.114  &   1.129    &  1.162\\
    \multicolumn{10}{l}{$\beta_2$ (Time-varying covariates)}  \\ 
    & Outpatient visit\footnotemark[1] & -0.184 & -0.334 & -0.521 & -0.608 & -0.654 & -0.642 &   -0.649    & -0.710  \\
    &New hospitalizations\footnotemark[1] & -0.399 & -0.301 & -0.333 & -0.331 & -0.318 & -0.315 &   -0.258    &  -0.235 \\
    &New organ failure events\footnotemark[1] & 0.048  & -0.018 & -0.021 & -0.020 & -0.019 & -0.019 &  -0.021     & -0.016 \\
    &No. of new comorbidities & 0.004  & 0.079  & 0.050  & 0.074  & 0.092  & 0.090  &   0.068    & 0.059 \\
    &No. of recent operations & 0.222  & 0.046  & -0.037 & -0.092 & -0.108 & -0.104 &   -0.148    & -0.232 \\
    &Change in labs & -0.002  & -0.002 &  -0.001 & -0.001 &  -0.001 &  -0.001 &      -0.001 & -0.001  \\ 
        $\rho$   & &-0.339 & -0.206 & -0.130 & -0.095  & -0.042 & -0.046 & -0.017 & 0.011 \\
    $\delta_0$& & 1.097   & 0.760  & 0.878  & 0.924  & 0.988  & 1.008     & 1.045   & 1.098 \\
    $\delta_1$ & & -0.053 & -0.023 & -0.025 & -0.015 & -0.022 & -0.025 &  -0.021 & -0.014 \\ \hline
    \end{tabular}%
\end{table}%
\footnotetext[1]{Indicates binary variables.}
	    \footnotetext[2]{Measured during one-year period prior to PH diagnosis.}
\end{landscape}

\subsection{Comparison with inferences based on risk set matching}
 
Risk set matching (RSM) serves as a useful reference for examining the relative performance of our proposed approach.
Table~\ref{tab:rsm} presents the one-year post-indication survival rates within the treatment and control groups, where controls are identified using risk-set matching. 
Here, for each treated patient with observed indication time $t$, we identified the not-yet-treated patient that was most similar in terms of their probability of receiving treatment based on covariates observed up to time $t$.

\begin{table}[htbp]
	\centering
	\caption{One-year survival rates following treatment indication for treated patients and controls identified using risk set matching}
	\label{tab:rsm}%
	\begin{tabular}{lccc}
		\hline
		\multirow{1.5}{*}{Study} & \multicolumn{2}{c}{One-year survival}& \multicolumn{1}{c}{Estimated}\\
		 \multirow{1.5}{*}{period} & \multicolumn{2}{c}{after indication} & \multicolumn{1}{c}{impact}\\ \cline{2-3}
		& Treatment & Matched Control & of PDE5I \\ \hline
		14 Days & 71.5\% & 71.5\% & \text{ }0.0\%  \\
		30 Days & 73.2\% & 73.6\% & -0.5\% \\
		60 Days & 75.1\% & 76.7\% & -1.6\% \\
		90 Days & 72.5\% & 76.9\% & -4.4\% \\
		120 Days & 73.0\% & 77.4\% & -4.4\% \\
		180 Days & 73.6\% & 77.7\% &  -4.1\% \\
		270 Days & 74.0\% & 77.1\% & -3.3\% \\
		365 Days & 72.5\% & 75.8\% & -3.1\% \\
				\hline
	\end{tabular}%
\end{table}%

These findings generally agree with the results previously described. 
In general, we find that estimates obtained using RSM are systematically smaller in magnitude than results based on the proposed approach.
However, because these estimates reflect the average effect of treatment on the treated rather than the average treatment effect across the entire patient population, these differences are not surprising. 
In particular, estimates calculated using RSM reflect the effects of \textit{delaying} treatment with PDE5Is for the subset of patients that have indications for treatment during the study period.
This means that for each study period, patients treated with PDE5Is during that period are compared to a matched control group consisting of patients who had not yet received treatment at that point but may have received treatment at a later point in the study. 
The main differences between the two approaches therefore lays in the interpretation of the parameters. 
Where the model proposed in Section~\ref{sec:model} yields an estimate that captures the impact of assignment to treatment versus control at time $t$ on one-year survival for all individuals in the population who had indications for treatment at that time, inferences based on RSM yield an estimate of the one-year survival rate for individuals who were indicated versus not yet indicated for treatment at time $t$ given the confounders measured up to that time.

\section{Discussion}
\label{sec:discuss}
In this paper, we propose a novel conceptualization of longitudinal observational studies with treatment by indication and provide a template for the design and analysis of such studies in a manner that approximates a randomized experiment with a binary treatment. 
This conceptualization reformulates the problem of evaluating a time-varying treatment in the presence of time-varying confounders as one of evaluating a binary treatment administered at a fixed time point, where the time of assignment is a partially observed pre-treatment confounder.
Our hope is that this simplified representation of a traditionally complex data structure will allow for more straightforward analyses of health data in the digital age (e.g., data from electronic medical records). 

The merit of the proposed model is that it allows for model-based assessments of the times of treatment indication based on disease progression that can accommodate time-varying covariates measured at intermittent observation times (i.e., when there is missing data and/or substantial variation in the timing of patient-provider visits). 
Our approach allows for systematic evaluation of the underlying health factors that may be most influential in determining a patient's need for treatment. 
Inferences based on this modeling strategy may be useful to address a number of important questions in health services research - for example, what preventative health strategies might be most effective for delaying the onset of indications for treatment? 
The model also allows us to obtain conditional probabilities for a subject having indications for treatment at different points over time, which could be used to inform preventative treatment strategies. We note that the proposed model assumes that subjects' overall health fluctuates under natural conditions.

As described in Section~\ref{sec:model}, the proposed model for time of indication for treatment can accommodate both fixed and time-varying covariates, which can be useful in explaining differences in the aspects of health associated with subject-specific characteristics and/or conditions that vary between hospital visit within a subject. 
Alternatively, covariate data could be excluded entirely from the model to make inferences about indication times that are viewed as fully stochastic. 
Another possibility for modeling variability in the parameters involves the inclusion of provider-level, or geographic-specific random effects. However, this complicates the evaluation of the likelihood, rendering model fitting more challenging.

Finally, we note that our application and inferential results should be regarded as an illustrative example of the proposed methods rather than an attempt to provide definitive answers about the causal effects of inappropriate prescribing practices on health-related outcomes for PH patients. 
In particular, for researchers interested in drawing causal inferences from observational studies with time-varying exposures, which often require a complex and challenging design process that can obfuscate any resulting inferences, the conceptual framework we present offers an alternative formulation of the underlying causal problem that is both intuitive and relatively straightforward to implement. 
The proposed approach can also be flexibly extended to accommodate a range of data structures, for instance in studies with multivariate outcomes.

We believe the framework presented in this paper will help expand the scope and usability of data from observational longitudinal studies and will facilitate investigation of these rich data sources across a wide variety of disciplines.
While our approach was developed in a medical setting, the applicability is far more general and offers natural extensions for applied research in education, psychopathology, and social science \citep[see, e.g.,][for a discussion of several longitudinal observational studies in these domains]{willett1998design}.
For instance, the methods described here could be used to engage with a longstanding debate in education research concerning the impact of parental divorce on educational achievement for children (e.g., \citealp{ayoub1999emotional, kim2011consequences, brand2019does}). 
When we define the individual-level effect of a divorce on educational attainment, we are faced with a seemingly time-varying treatment (i.e., a parental divorce can occur at many points in time throughout childhood) and a time-to-event outcome (e.g., time to high school completion) \citep{brand200711}. 
Under the proposed framework, the time of indication for divorce could be conceptualized as the first-hitting-time of a latent state process that represents marital stability. 
By inferring the plausible times of indication for divorce for families who remained married during a given study period, our approach could be used to obtain more precise estimates of the effects of the decision to divorce on time to degree completion for individuals who experience family stress and/or marital instability during childhood \citep{sun2002children}.
Our framework could also be applied to study the causal effects of the decision to attend or stop attending a religious service on the time to remission for depression \citep{vanderweele2016causal}, or to estimate the effects of income loss on all-cause mortality \citep{etches2009economic}. 
The wide applicability of our approach, as illustrated by these several example observational studies, can help advance our collective understanding of best practices in this important domain.

\appendix

\section*{Acknowledgements}
This work was supported in part by the VA HSR\&D IIR 15-115.

\section{Covariate selection procedure for VA application}
\label{app:covs}

Covariate measurements were collected for each patient during each visit that occurred in the study period, and each of the 534 treated patients in our analysis was observed for between one to 611 distinct visits during the study period. For variable selection, we first constructed a new dataset containing observations for each of the 534 treated patients at each of two time periods: 1) the visit associated with assignment to treatment, and 2) the preceding visit. Patients who received assignment to treatment at their first visit were included in the dataset only once.  Each observation vector in the resulting dataset therefore corresponds to the covariate values of a patient who has not yet been assigned or to a patient at their time of assignment. In principle, contrasts of covariates between these two groups should capture information about how the latent health process - and the corresponding indication for assignment to either treatment or control -  varies with these longitudinal measurements. 

For variable selection, we performed random forest analysis \citep{breiman2001random} implemented using the ``randomForest" package in R \citep{breiman2003manual} to evaluate the relative importance of each of the time-varying covariates for predicting the time of indication for treatment. Using this procedure, we identified the 20 most influential time-varying covariates from over 150 available variables. Among these 20 covariates were several related to the current physical characteristics of the patient (e.g., current weight and blood pressure) as well as variables that captured changes in these physical measurements (e.g., change in weight). Other important covariates include the number of comorbidities that the patient has at the time of the visit and changes in health care utilization such as recent organ failure events (e.g., right heart failure), recent inpatient visits, or receipt of incidental procedures (e.g., echocardiograph).

\newpage
\singlespacing
\bibliographystyle{apalike}
\bibliography{pde5iREF}
\doublespacing

\end{document}